\documentclass[12pt]{article}
\title{Phase Diagrams of the Spin-1 Blume-Capel Film With an Alternating Crystal Field}
\date{September 26, 2003}
\author{Hamid EZ-ZAHRAOUY$^{(a)} $\footnote{Corresponding author: ezahamid@fsr.ac.ma} and Ahmed KASSOU-OU-ALI$^{(b)}$}

\begin{document}
\maketitle
\begin{center}
{\it \small
(a) Facult\'e des Sciences, D\'epartement de Physique, Laboratoire de Magn\'etisme et Physique des Hautes Energies, B.P. 1014, Rabat, Morocco.\\
(b) Facult\'e des Sciences, D\'epartement de Physique, Laboratoire de Physique Th\'eorique, B.P. 1014, Rabat, Morocco.}
\end{center}

\abstract{The spin-1 ferromagnetic Blume-Capel film with an alternating crystal field $\Delta=\Delta_1$ on the odd layers and $\Delta=\Delta_2$ on the even ones is considered in the mean field approximation. The ground state phase diagrams in the ($d_1,d_2$) plane ($d_1=\Delta_1/J$ and $d_2=\Delta_2/J$) are determined analytically; the number of their phases depends on the parity of the number of layers of the film. At finite temperature, fifteen types of topology, depending on the range of variation of $d_1$, are found in the (t,$d_2$) plane for an even number of layers, but only fourteen for an odd number of layers. The phase diagrams exhibit a variety of multicritical points. In particular, a tricritical point C appears in the paramagnetic-ferromagnetic line of transition, but only for values of $d_1$ larger than a threshold value $d_{trc}^{(2)} ={8 \ln 2 \over 3}$ which is but the tricritical crystal field (in the mean field approximation) of the spin-1 Blume-Capel model on a square lattice. Moreover, lines of transition presenting the reentrant and double reentrant behaviors may also take place.}

\newpage
\section{Introduction}
\setlength{\parskip}{.2in} 
The Blume-Capel model is a spin-1 Ising model with single-ion anisotropy. It has been originally introduced to study first-order magnetic phase transitions [1,2] and then applied to multicomponent fluids [3]. Later, it was generalized to the Blume-Emery-Griffiths model [4] to study the He$^3$-He$^3$  mixtures and a variety of other physical systems.
   
The BC model has been investigated in detail using many approximate methods, namely mean field approximation [1,2], high temperature series expansion [5], constant-coupling approximation [6], Monte-Carlo [7] and renormalization-group [8] techniques. All of these approximate schemes suggest the existence of a tricritical point at which the system changes from the second-order phase transition to the first-order one, when the transition temperature is plotted as a function of the parameter of anisotropy.

On the other hand the BC model has been investigated on semi-infinite lattices with modified surface coupling. Several approximate methods have been used [9] (and references therein). All of them show the possibility to have a phase with ordered surface and disordered bulk, and show the existence of the so-called 'extraordinary', 'surface' and 'ordinary' transitions, and of multicritical points called special points [10]. More recently, with the development of the molecular beam epitaxy technique and its application to the growth of thin metallic films, renewed interest in thin film magnetism has been stimulated, and much experimental and theoretical efforts have been devoted to the subject [11]. The BC model with a rationally decreasing crystal field has recently been extended to a layered structure and analyzed in the mean field approximation [12]. The model exhibits a constant tricritical point and a reentrant phenomenon for a certain number of layers.

Our aim in this work is to study, using mean field theory, the phase diagrams of a ferromagnetic spin-1 Blume-Capel film with an alternating crystal field: $\Delta=\Delta_1$ on the odd layers and $\Delta=\Delta_2$ on the even ones. The number of topologies of the phase diagrams is finite and depends only on the parity of the number of layers of the film. The phase diagrams present many multicritical points and may have transition lines exhibiting the reentrant and double-reentrant behaviors; this kind of behavior has been previously found in  the Paramagnetic-Ferromagnetic transition line of the ordinary Blume-Capel model with a random crystal field [13] or with random nearest-neighbor interactions[14].         
 
The paper is organized as follows. In section 2 we present the model and give the mean field equations of the different order parameters. In section 3 we present the results, then we end in section 4 by a conclusion.       

\section{Model and method}
The system we consider is the spin-1 Blume-Capel model of a film of N ferromagnetically coupled layers with an alternating crystal field. The Hamiltonian of this system is given by
\begin{equation}                                 
{\cal H}=-J\sum_{\langle i,j \rangle}S_iS_j+\sum_i \Delta_i S_i^2  
\end{equation}                                   
where $S_i=\pm1,0$; the first sum runs over all pairs of nearest-neighbors and $\Delta_i$ is the crystal field which takes the value $\Delta_1$ on the odd layers and $\Delta_2$ on the even ones. 
The mean field equations of state are straightforwardly obtained and are given by 
\begin{equation}                                 
m_k=\langle S_k \rangle={2\sinh \beta h_k \over e^{\beta \Delta_k}+2 \cosh \beta h_k}
\label{mk}
\end{equation}                                   
\begin{equation}                                 
q_k=\langle S_k^2 \rangle={2\cosh \beta h_k \over e^{\beta \Delta_k}+2 \cosh \beta h_k}
\end{equation}                                   
with 
\begin{equation}
\begin{array}{l}                                      
h_1=J(zm_1+m_2)\\
h_k=J(m_{k-1}+zm_k+m_{k+1}) \makebox{, for } 2 \leq k \leq N-1\\
h_N=J(m_{N-1}+zm_N)
\end{array}     
\end{equation}                                 

In these equations, $m_k$ ($q_k$) is the reduced magnetization (quadrupolar moment) of the   layer;  and $z=4$ is the interlayer coordination number. The reduced total magnetization is $m={1 \over N}\sum_{k=1}^N m_k$. The reduced free energy of the film is given by
\begin{eqnarray}
f=-{1\over \beta} \sum_{k=1}^N \ln \left[ 1+2e^{-\beta \Delta_k} \cosh \beta h_k \right]+{1 \over 2}\sum_{k=2}^{N-1}(m_{k-1}+zm_k+m_{k+1})m_k \nonumber \\
+{1 \over 2}m_1(4m_1+m_2)+{1 \over 2}m_N(m_{N-1}+4m_N)
\label{fren}
\end{eqnarray}                                 

\section{Results and discussions}

The study of the phase diagrams in the (t,$d_2$) plane exhibits the existence of different topologies depending on the range of variation of $d_1$ ($d_1=\Delta_1/J$, $d_2=\Delta_2/J$ and $t=T/J$ are the values of the reduced crystal fields and temperature respectively). We note from the beginning that these topologies are limited in number and don't depend on the number of layers itself but only on its parity. In the following we will discuss in detail the case of an even number of layers and comment at the end on the odd number case. To understand the origin of the different topologies, we first determine analytically the ground state phase diagram in the ($d_1,d_2$) plane by looking for the lowest energy configurations. For an arbitrary even number of layers $N=2L$, we distinguish six regions of this plane (Fig.1(a)) (but only five for an odd number of layers (Fig.1(b))):
\begin{itemize}
\item[(a)] $d_1<3$, $d_2<3$ and $d_1+d_2<6-1/L$; the stable configuration\footnote[1]{We use obvious notations for the model where the natural basic entity of the different states and phases is a bilayer} is $(11)^L$, where the spins of the sites of all the layers are all equal to 1.
\item[(b)] $3<d_1<4$, $d_2<d_1$ and $(L-1)d_1+Ld_2<6L-4$; the stable configuration is $01(11)^{L-1}$, where the spins of the first layer are all equal to 0 while those of the other layers are all equal to 1.
\item[(c)] $d_1>4$ and $d_2<2$; the stable configuration is $(01)^L$ where the spins of the odd layers are all equal to 0 while those of the even ones are all equal to 1.
\item[(d)] $3<d_2<4$, $d_1<d_2$ and $Ld_1+(L-1)d_2<6L-4$; the stable configuration is $(11)^{L-1}10$ .
\item[(e)] $d_1<2$ and $d_2>4$; the stable configuration is $(10)^L$. 
\item[(f)] $d_1>2$, $d_2>2$, $d_1+d_2>6-1/L$, $(L-1)d_1+Ld_2>6L-4$ and $Ld_1+(L-1)d_2>6L-4$; the stable configuration is the non magnetic state $(00)^L$.	
\end{itemize}

From Fig.1(a) it is clear that for fixed values of $d_1$, the system exhibits various transitions by varying $d_2$. For the following ranges of variation of $d_1$: i) $0<d_1<2$, ii) $2<d_1<3-1/L$, iii) $3-1/L<d_1<3$, iv) $3<d_1<4$ and v) $d_1>4$, the system presents the following phase transitions respectively:
\begin{itemize}
\item[i)] from the state $(11)^L$ to the state $(11)^{L-1}10$ at $d_2=3$ and then to the state $(00)^L$ at $d_2=4$.
\item[ii)] from the state $(11)^L$ to the state $(11)^{L-1}10$ at $d_2=3$ and then to the state $(00)^L$ at $Ld_2=(6L-4)-(L-1)d_1$.
\item[iii)] from the state $(11)^L$ to the state $(00)^L$ at $d_2=(6-1/L)-d_1$.
\item[iv)] from the state $01(11)^{L-1}$ to the state $(00)^L$ at $(L-1)d_2=(6L-4)-Ld_1$.
\item[v)] from the state $(01)^L$ to the state $(00)^L$ at $d_2=2$.   
\end{itemize}

It is worth to note here that, contrary to the two or three dimensional Blume-Capel cases (or the case of a film with a homogenous crystal field), the ground state of certain layers may be magnetic even for relatively large values of the crystal field in these layers. For example, for $d_1<2$, the even layers (except the final one) are in the magnetic states $m_{2k}=1$ even with a crystal field, in these layers, reaching the value $d_2=4$ (which is larger than the critical crystal field for a square ($d_c^{(2)}$=2) or three dimensional cubic ($d_c^{(3)}$=3) lattices).

 For finite temperatures the transition lines are determined by solving numerically the equations~\ref{mk} which may have more than one solution; the one which minimizes the free energy~\ref{fren} corresponds to the stable phase. Transitions of the second order are characterized by a continuous vanishing of the magnetizations while those of the first order exhibit discontinuities at the transition points.

In order to describe the different entities in the phase diagrams, the Griffiths notations [15] will be adopted: the critical end-point $B^mA^n$ denotes the intersection of $m$ lines of second order and $n$ of first order; the multicritical point $C$  denotes the intersection of m lines of second order; the tricritical point, which is the intersection of a line of second order and a line of first order, is denoted in particular by $C$. In addition, the following notations of the different phases will be adopted: $(FF)^L$ is, for example, a state where all the layers are in the ferromagnetic state, i.e. the magnetization and the quadrupolar moment of each layer are both different from zero. $(FF)^{L-1}F0$ is a phase where all the layers are in the ferromagnetic state except the last one which is in the non magnetic state (noted by 0) and is characterized by  vanishing magnetization and quadrupolar moment. $(PP)^L$ is a phase where all the layers are in the paramagnetic state i.e. the magnetization of each layer is equal to zero but not the quadrupolar moment…     

For an even number of layers ($N=2L$), fifteen types of topology of the phase diagram in the (t,$d_2$) plane are found. They may be gathered into six groups, each one is characterized by a certain range of variation of $d_1$.

1) For $0<d_1<d_{trc}^{(2)}$ (e.g. Fig.2(a)), there is a critical line separating the paramagnetic phase $(PP)^L$ from the ferromagnetic ones $(FF)^L$, $(FF)^{L-1}F0$ and $(F0)^L$. These ferromagnetic phases are separated by two transition lines of  first-order at low temperatures which meet two critical lines at two tricritical points (totally inside the ordered phases) $C_1$ and $C_2$ respectively. The first-order lines reach the $d_2$ axis at the fixed values (independent of L) $d_2=3$ and $d_2=4$ respectively. The critical lines present both a reentrant behavior near the paramagnetic-ferromagnetic (P-F) transition line and meet this last at two multicritical points $(B^3)_1$ and $(B^3)_2$ respectively. The topology described here is characterized, contrary to the ordinary Blume-Capel model, by the absence of a tricritical point in the (P-F) transition line . The value of $d_1$ over which this point appears is found to be $d_{trc}^{(2)}\approx 1.848\approx {8 \ln 2 \over 3}$ which is but the mean field tricritical crystal field of the spin-1 Blume-Capel model on a square lattice.

2) For $d_{trc}^{(2)}<d_1<2$, four types of topology are found with increasing $d_1$, all of them present a tricritical point in the P-F transition line and present the same features of the type-1 topology with some differences. For $d_1$ in the vicinity of $d_{trc}^{(2)}$ the lines of transition (F-F lines) separating the ferromagnetic phases $(FF)^L$, $(FF)^{L-1}F0$ and $(F0)^L$ still exhibit the reentrant phenomenon near the P-F transition line and meet this one at the multicritical points $(B^3)_1$ and $(B^3)_2$ (e.g. Fig2(b)). With increasing $d_1$, the F-F lines meet the P-F line at the multicritical point $B^3$ and at the critical end-point $BA^2$ respectively,  and still present the reentrant behavior (e.g. Fig.2(c)); for higher values of $d_1$, the former lines lose the reentrant behavior and meet the P-F line at two distinct critical end points $(BA^2)_1$ and $(BA^2)_2$ as is depicted in Fig.2(d). Finally, for $d_1$ close to 2 the P-F first-order line develops a reentrant topology (e.g. Fig.2(e)).

Note that for $d_1<2$ the ferromagnetic phase $(F0)^L$ is stable for an infinitely large value of $d_2$ at temperatures lower than a temperature $t_c^{(2)}(d_1)$ which is independent of the number of layers $2L$. $t_c^{(2)}(d_1)$ is but the critical temperature of the ordinary two-dimensional spin-1 Blume-Capel model corresponding to $d_1$. This may easily be understood even beyond the mean field approximation. In fact at low temperatures and for large values of $d_2$, the odd layers are all non magnetic ($m_{2k+1}=0$) and the system of equations of the magnetizations $m_{2k}$ of the even layers is decoupled to L equivalent two-dimensional equations with a crystal field $d_1$; the magnetizations $m_{2k}$ (and then the total magnetization) vanish at the same temperature which is $t_c^{(2)}(d_1)$. For $d_1>2$ the ferromagnetic phase is unstable for large values of $d_2$, as will be seen, in accordance with the ground-state phase diagram.

3) For $2<d_1<3-1/L$, three types of topology exist. For these topologies the ferromagnetic state $(F0)^L$ disappears from the ordered phases. The ferromagnetic phases $(FF)^L$ and $(FF)^{L-1}F0$ are, as above, separated by a transition line of first-order, at low temperatures, which reaches the $d_2$ axis at $d_2=3$ and meets the second order line at a tricritical point $C_1$. This last line meets the P-F line at a critical end-point $BA^2$. For $d_1$ in the close vicinity of 2, the P-F line of transition exhibits a reentrant and double-reentrant behaviors (e.g. Fig.2(f)); with increasing $d_1$, only the double reentrant still exist (e.g. Fig.2(g)) and then it disappears for higher values of $d_1$ (e.g. Fig.2(h)). Note that the P-F transition line reaches, for $2<d_1<3-1/L$, the $d_2$ axis at $d_2$ given by $(L-1)d_2=(6L-4)-Ld_1$; the surface of the $(FF)^{L-1}F0$ phase then decreases with increasing $d_1$ and it collapses at $d_1=3-1/L$.

4) For $3-1/L<d_1<3$, the phase diagram is of the same type as that of the ordinary spin-1 Blume-Capel model (e.g. Fig.2(i)); the first-order line reaches the $d_2$ axis at $d_2$ given by $d_2=(6-1/L)-d_1$.

5) For $3<d_1<4$, three types of topology are found. These are characterized by the appearance of the ferromagnetic phase $0F(FF)^{L-1}$ at low temperatures. The ferromagnetic phases $(FF)^L$ and $0F(FF)^{L-1}$ are separated by a critical line of approximately constant temperature which meets the P-F transition line at a critical end-point $BA^2$. The P-F line reaches the $d_2$ axis at $d_2$ given by $Ld_2=(6L-4)-(L-1)d_1$. For $d_1$ far from 4, a typical phase diagram is given in Fig.2(j). With increasing $d_1$, the P-F transition line develops a double-reentrant behavior (e.g. Fig.2(k)), and when $d_1$ approaches 4 this line presents a reentrant and double-reentrant behaviors (e.g. Fig2(l)).

6) For $d_1>4$, three types of topology are fond; these are characterized by the appearance, in addition to the state $0F(FF)^{L-1}$, of the state $(0F)^L$ at low temperatures. The ferromagnetic phases $(FF)^L$, $0F(FF)^{L-1}$ and $(0F)^L$ are separated by two critical lines of approximately constant temperatures which meet the P-F transition line at two critical end points $(BA^2)_1$ and $(BA^2)_2$  respectively. The P-F line reaches the $d_2$ axis at $d_2=2$. For values of $d_1$ very close to 4 the P-F line presents a reentrance and a double reentrance (e.g. Fig.2(m)); the reentrance disappears with increasing $d_1$ (e.g. Fig.2(n)) and then the double reentrance disappears for high values of $d_1$ (e.g. Fig.2(o)).

For an odd number, $N=2L+1$, of layers, the ground-state (Fig.1(b)) has only five different phases (instead of six for N even); in fact, for small values of $d_1$ ($d_1<2$) only two phases now exist: $(11)^L1$ (for $d_2<4$) and $(10)^L1$ (for $d_2>4$).

For finite temperatures, fourteen types of topology exist in this case:
\begin{itemize}
\item[i)] For $0<d_1<d_{trc}^{(2)}$   the topology is like the type-1 above (in the case of an even number of layers) but with only two ferromagnetic phases $(FF)^LF$ and $(F0)^LF$ separated by a unique line having a tricritical point $C_1$ at low temperatures.
\item[ii)] For $d_{trc}^{(2)}<d_1<2$ there are four types of topology which  are like those of the group 2 above, but with the modification mentioned in i).
\item[iii)] For $2<d_1<3$, there are no topologies like those of the group 3 above. The topologies, in the number of three, are all of the ordinary Blume-Capel type, but for the two first ones ($d_1$ close to 2) the P-F transition line develops a reentrance for $d_1$ very close to 2, and a reentrance and double-reentrance for higher values of $d_1$ (Fig.3).
\item[iv)] For $3<d_1<4$, there are three types of topology which are like those of the group 5 above, but with the ferromagnetic phase $0F(FF)^{L-1}0$ at low temperatures (instead of $0F(FF)^{L-1}$ for N even).
\item[v)] For $d_1>4$, there are three topologies which are  like those of  the group 6 above, but with the ferromagnetic phases $(0F)^L0$ and $0F(FF)^{L-1}0$ at low temperatures (instead of $(0F)^L$ and $0F(FF)^{L-1}$  for N even).
\end{itemize}
	 
\section{Conclusion}

We have studied, using mean field theory, the phase diagrams of a ferromagnetic spin-1 Blume-Capel film with an alternating crystal field: $\Delta=\Delta_1$ on the odd layers and $\Delta=\Delta_2$ on the even ones. The ground state phase diagrams in the $(d_1,d_2)$ plane ($d_1=\Delta_1/J$ and $d_2=\Delta_2/J$) are determined analytically; the number of their phases depends on the parity of the number of layers of the film. At finite temperatures, fifteen types of topology, depending on the range of variation of $d_1$ , are found in the (t,$d_2$) plane for an even number of layers, but only fourteen for an odd number of layers. The phase diagrams exhibit many multicritical points. In particular, a tricritical point $C$ appears in the paramagnetic-ferromagnetic line of transition, but only for values of $d_1$ larger than a threshold value $d_{trc}^{(2)} ={8 \ln 2 \over 3}$ which is but the mean field tricritical crystal field of the spin-1 Blume-Capel model on a square lattice. Moreover, lines of transition presenting the reentrant and double-reentrant behaviors also appear in the phase diagrams. 

\newpage
\section*{References}

\begin{description}
\item[][1] M. Blume, Phys. Rev. {\bf 141}, 517 (1966)
\item[][2] H.W. Capel, Physica {\bf 32}, 966 (1966)
\item[][5] D.M. Saul and M. Wortis, Amer. Inst. Phys. Conf. Proc. {\bf 5}, 349 (1972)\\
    P.F. Fox and D.S. Gaunt, J. Phys. C {\bf 5}, 3085 (1972)\\
    D.M. Saul, M. Wortis and D. Stauffer, Phys. Rev. B {\bf 9}, 4964 (1974)
\item[][6] M. Tanaka and K. Takahashi, Phys. Stat. Sol. (b) {\bf 93}, K85 (1979)
\item[][7] B.L. Arora and D.P. Landau, Amer. Inst. Phys. Conf. Proc. {\bf 5}, 352 (1972)\\
    A.K. Jain and D.P. Landau, Phys. Rev. B {\bf 22}, 445 (1980)\\
    W. Selk and J. Yeomans, J. Phys. A {\bf 16}, 2789 (1983)\\
    C.M. Care, J. Phys. A {\bf 26}, 1481 (1993)\\
    M. Deserno, Phys. Rev. E {\bf 56}, 5204 (1997)
\item[][8] A.N. Berker and M. Wortis, Phys. Rev. B {\bf 14}, 4946 (1676)\\
      T.W. Bukhard, Phys. Rev. B {\bf 14}, 1196 (1976)\\
      H. Dickinson and J. Yeomas, J. Phys. C, L345 (1983)
\item[][9] C. Buzano and A. Pelizzola, cond-mat / 9502017 (1996)
\item[][10] A. Benyoussef, N. Boccara and M. Saber, J. Phys. C {\bf 19}, 1983 (1986)
\item[][11] A. Benyoussef and H. Ez-Zahraouy, Phys. Script. {\bf 57}, 603 (1998)
\item[][12] L. Bahmad, A. Benyoussef and H. Ez-Zahraouy, J. Magn. Magn. Mat. {\bf 251}, 115 (2002)
\item[][13] A. Maritan, M. Cieplak, M.R. Swift, F. Toigo and  J.R. Banavar, Phys. Rev. Let. {\bf 69} (1992) 221\\
A. Benyoussef, T. Biaz, M. Saber and M. Touzani, J.Phys. C: Solid State Phys. {\bf 20} (1987) 5349\\
T. Kaneyoshi and J. Mielnicki, J.Phys. C: Condens.Matter {\bf 2} (1990) 8773 
\item[][14] T. Kaneyoshi, J. Magn. Magn. Mat. {\bf 92} (1990) 59 
\item[][15] R.B. Griffiths, Phys. Rev. B {\bf 12}, 345 (1975)
\end{description}

\newpage
\section*{Figure captions}

\begin{description}
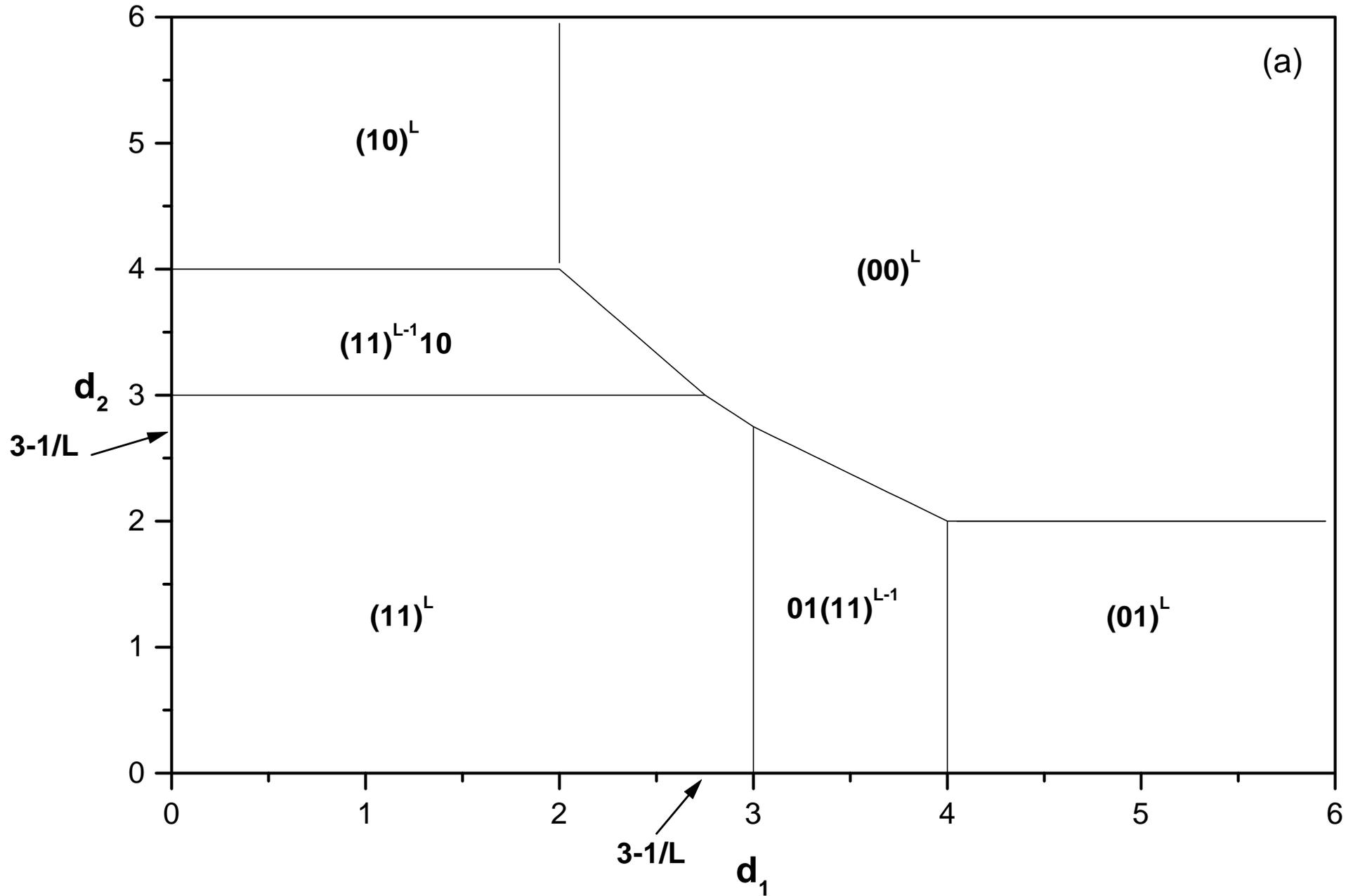
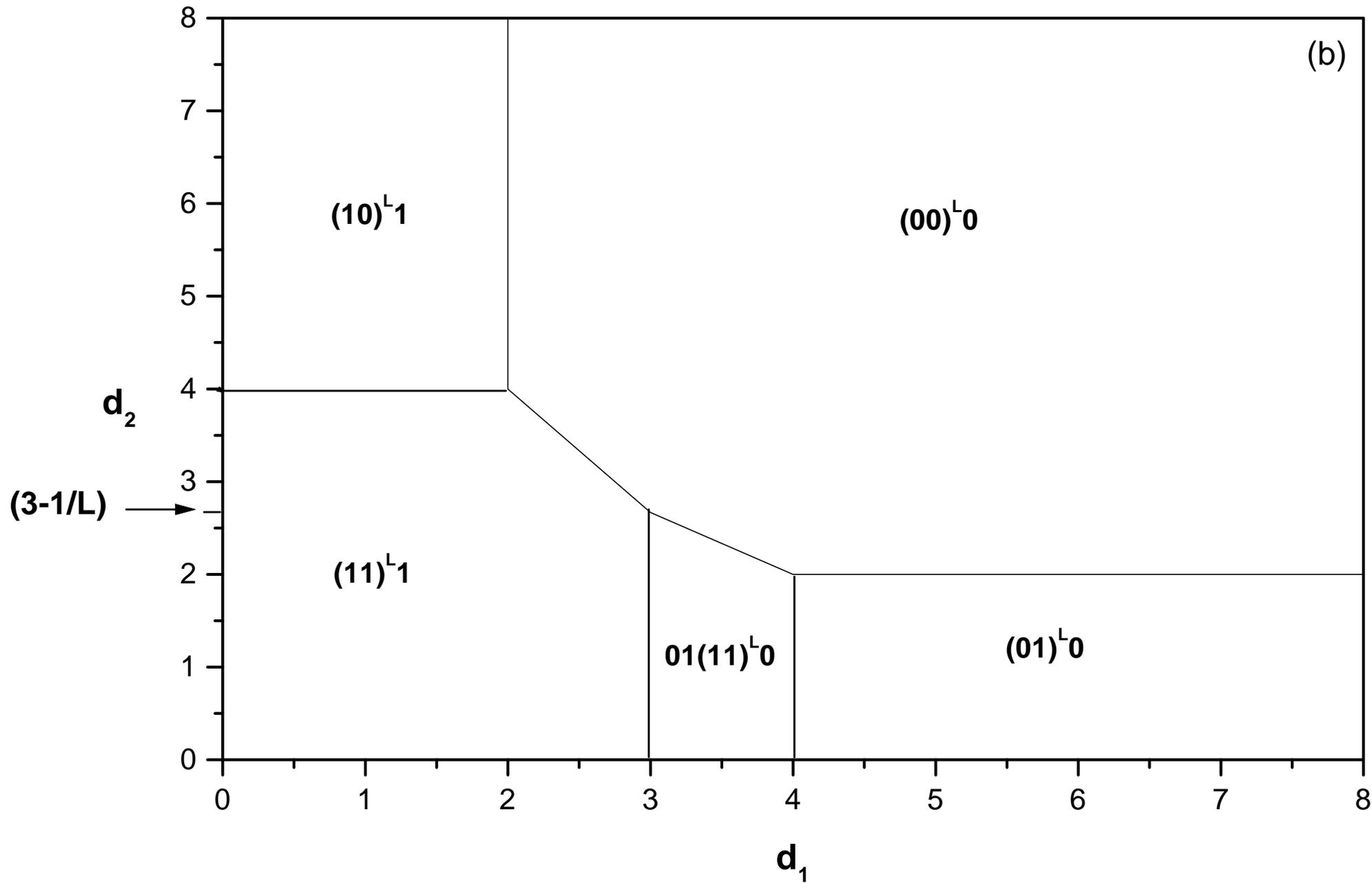
\item[Fig.1] The ground state phase diagrams in the $(d_1,d_2)$ plane for an arbitrary number N of layers, (a) $N=2L$, (b) $N=2L+1$. 
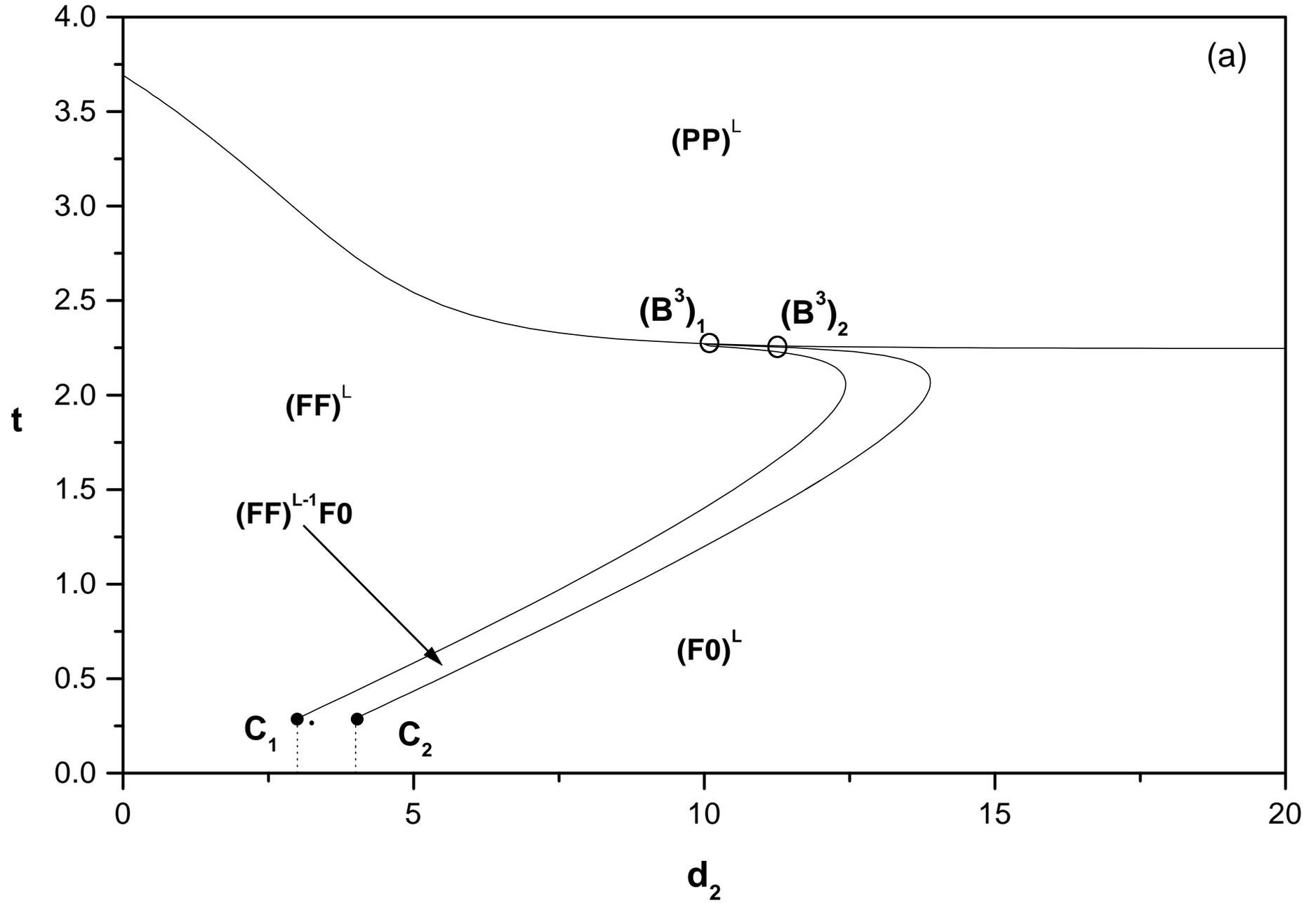
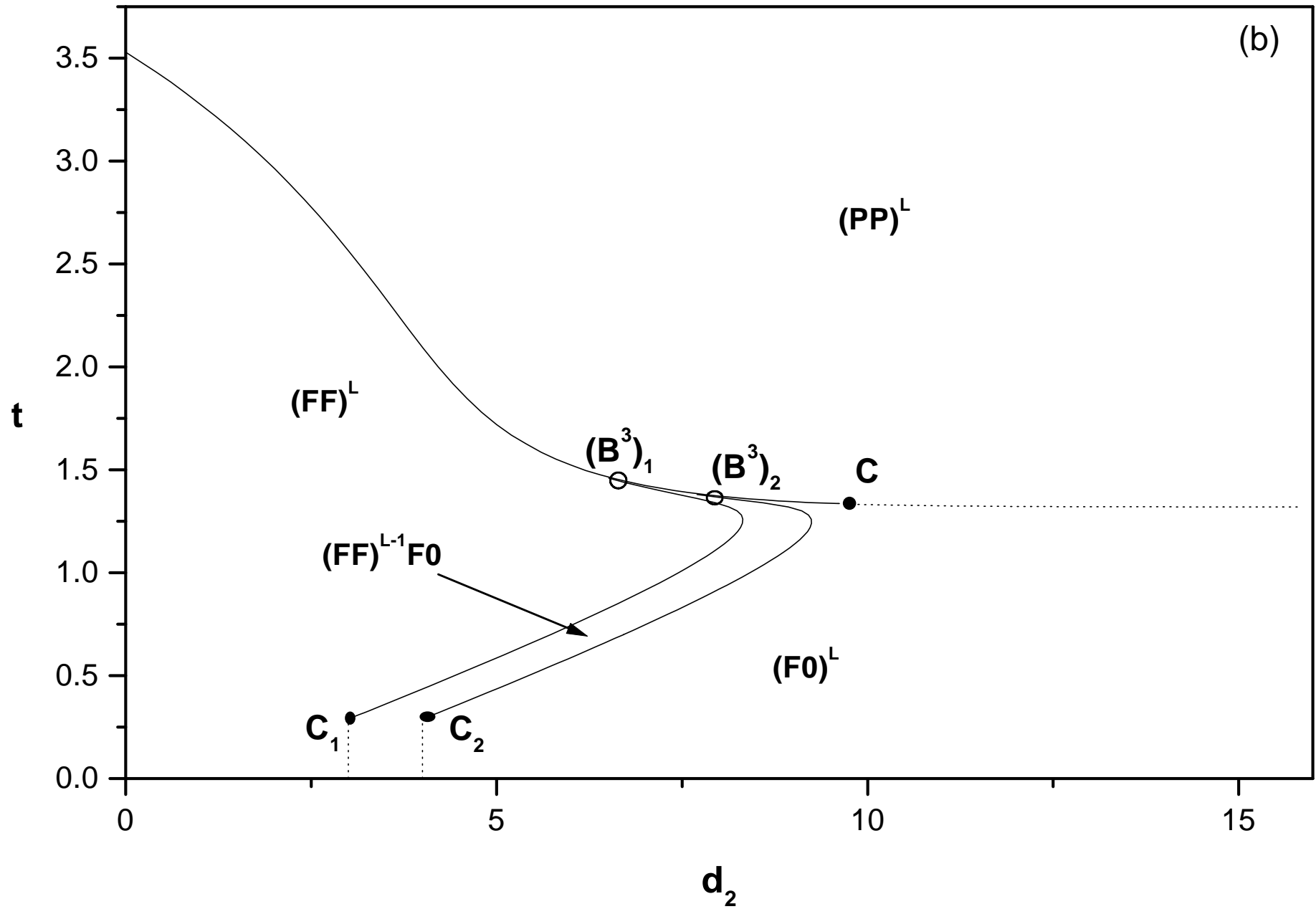
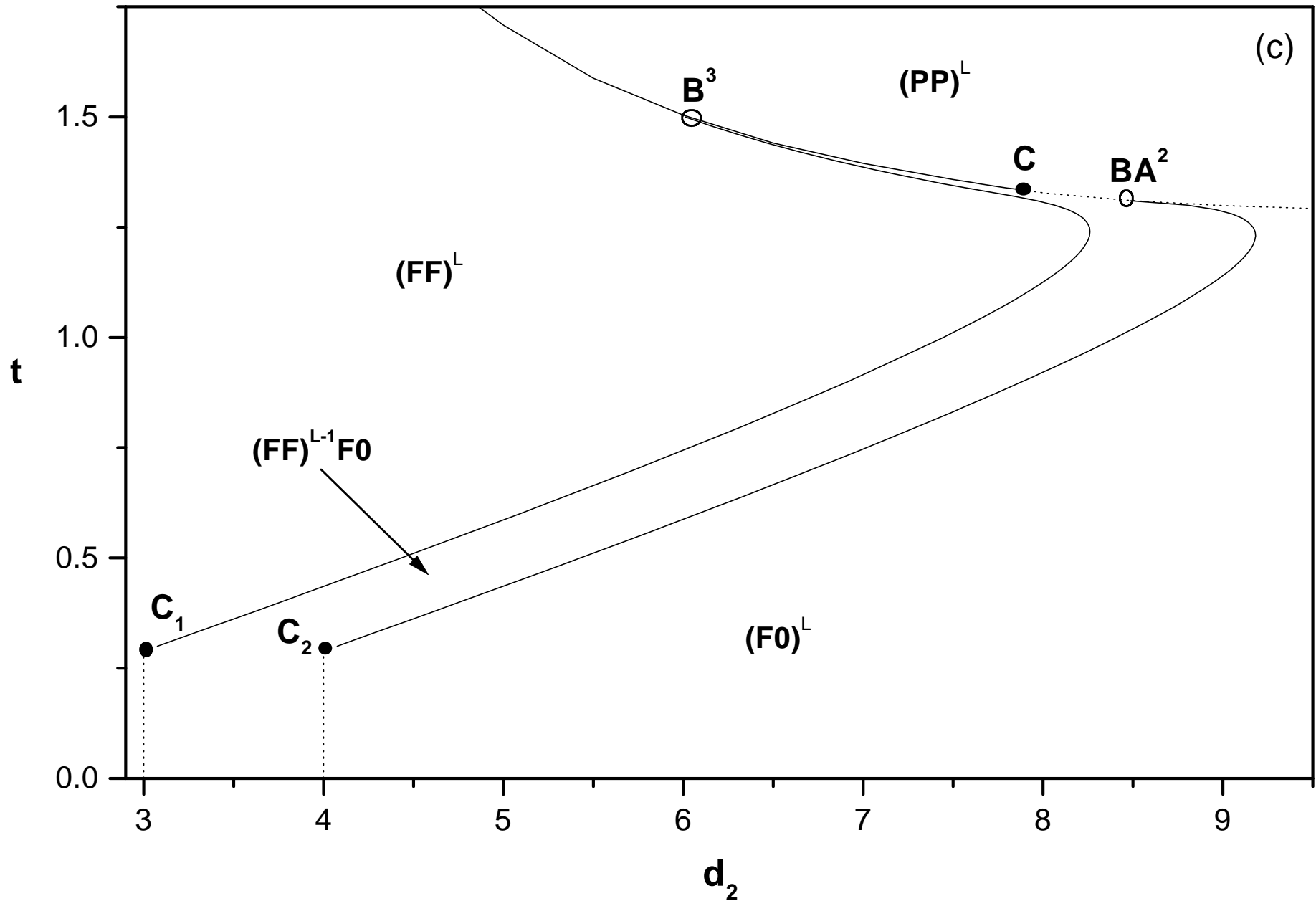
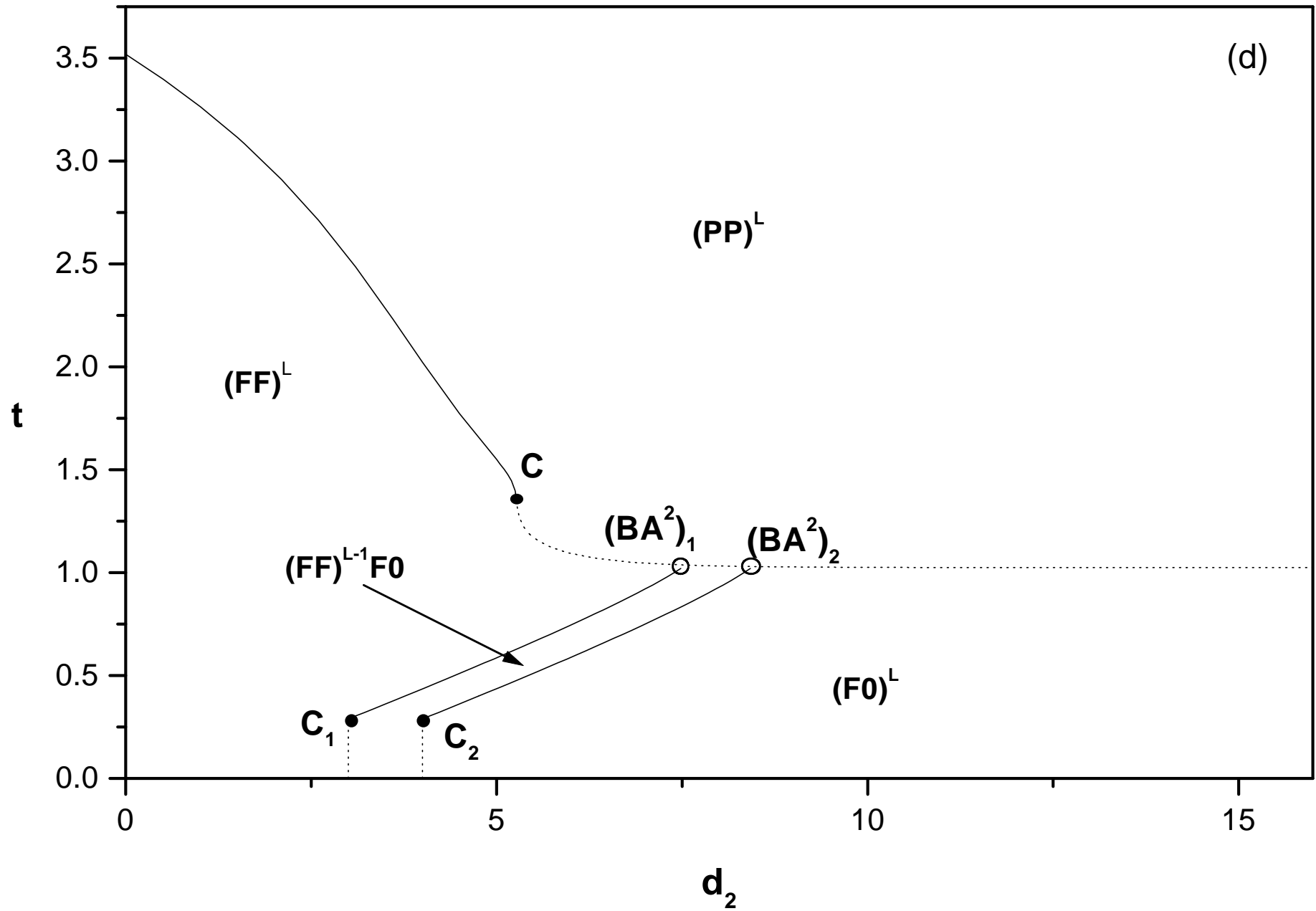
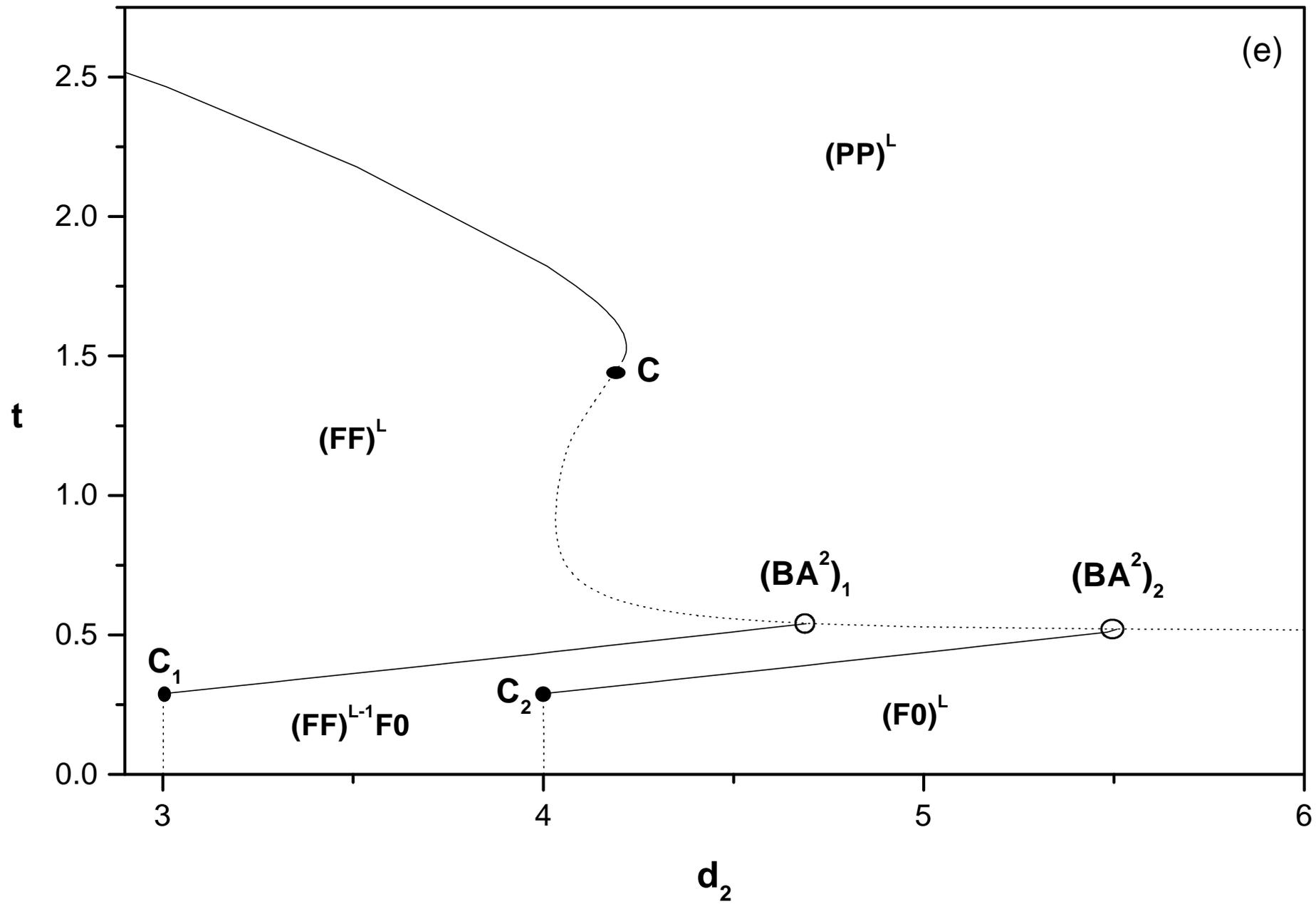
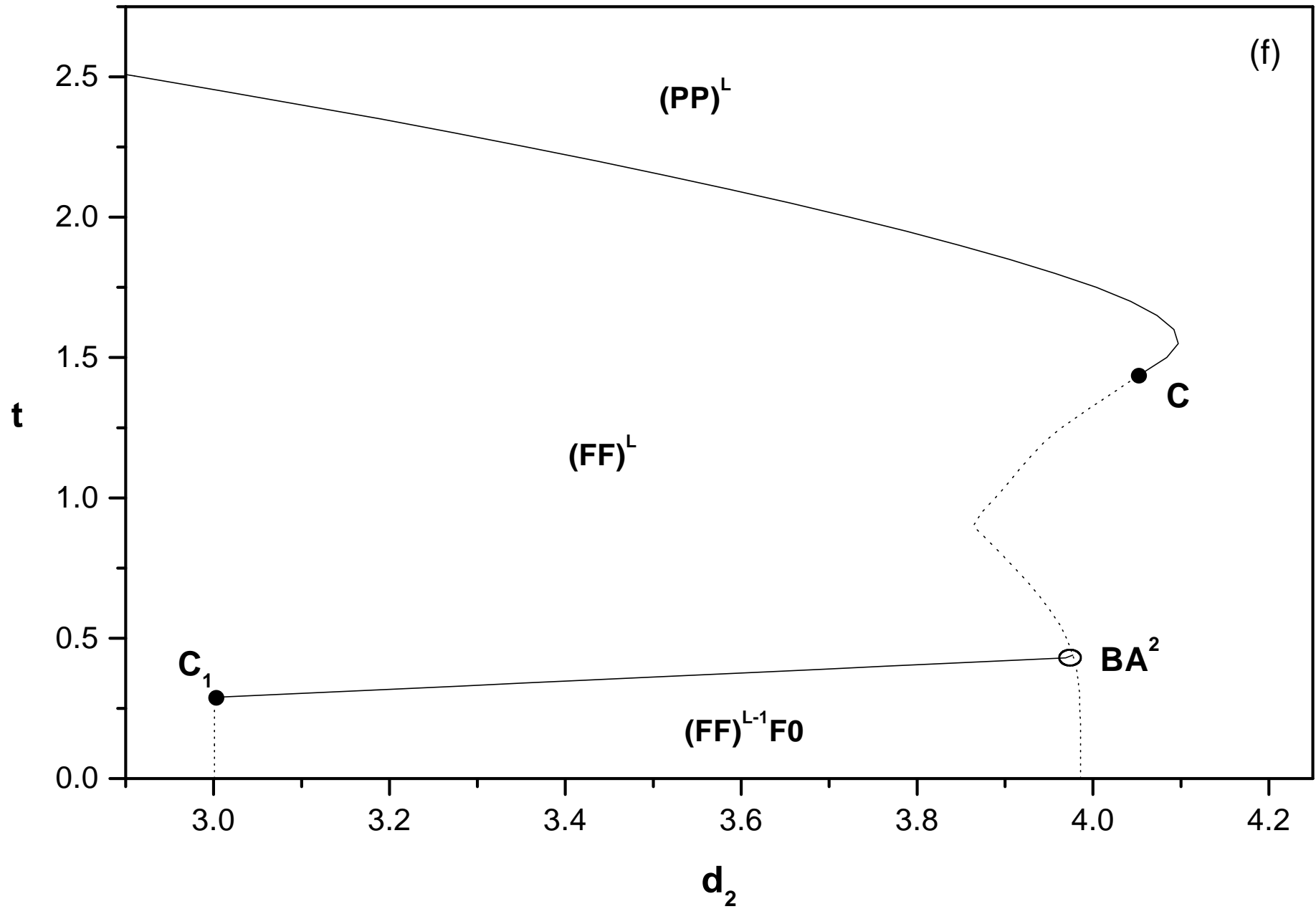
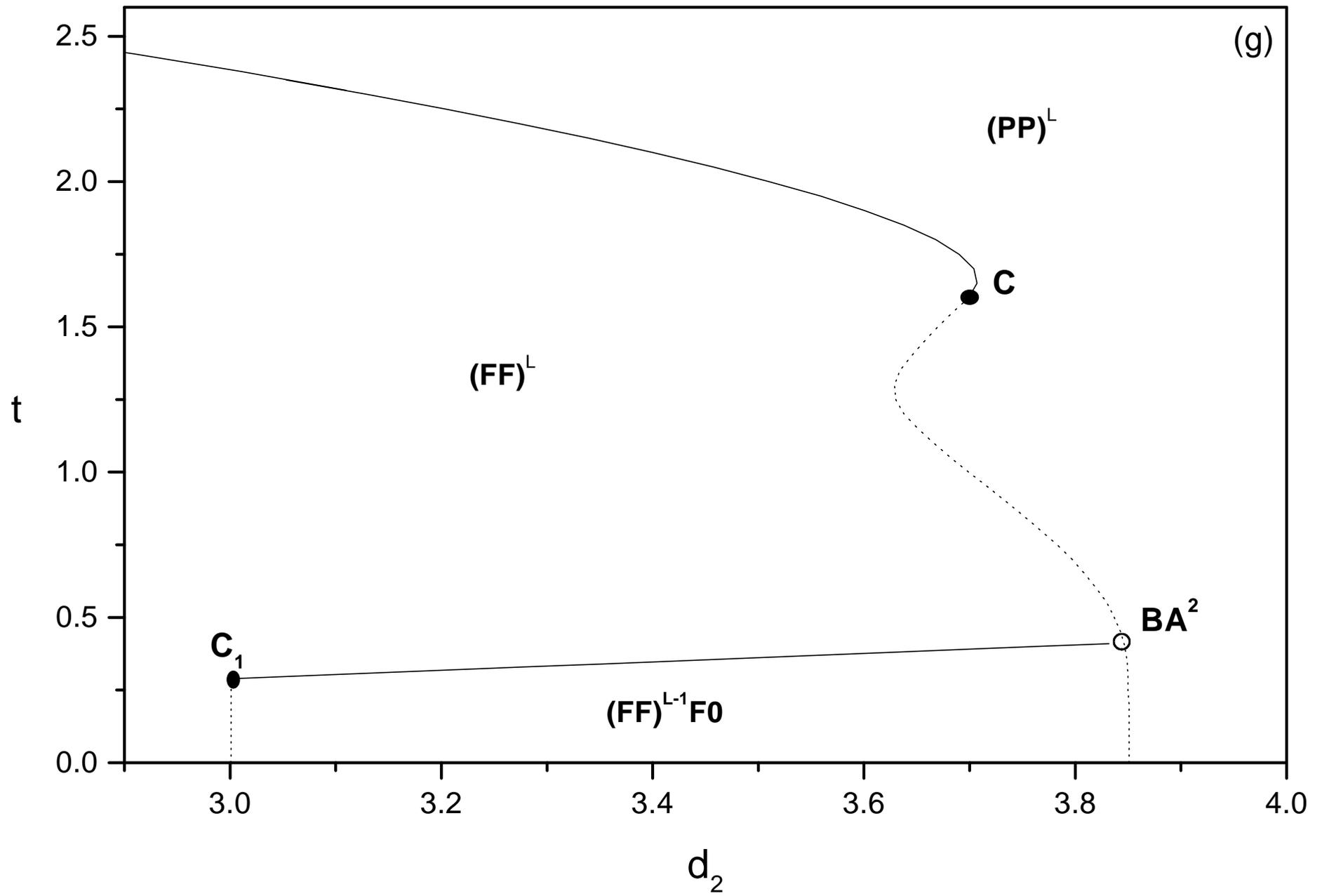
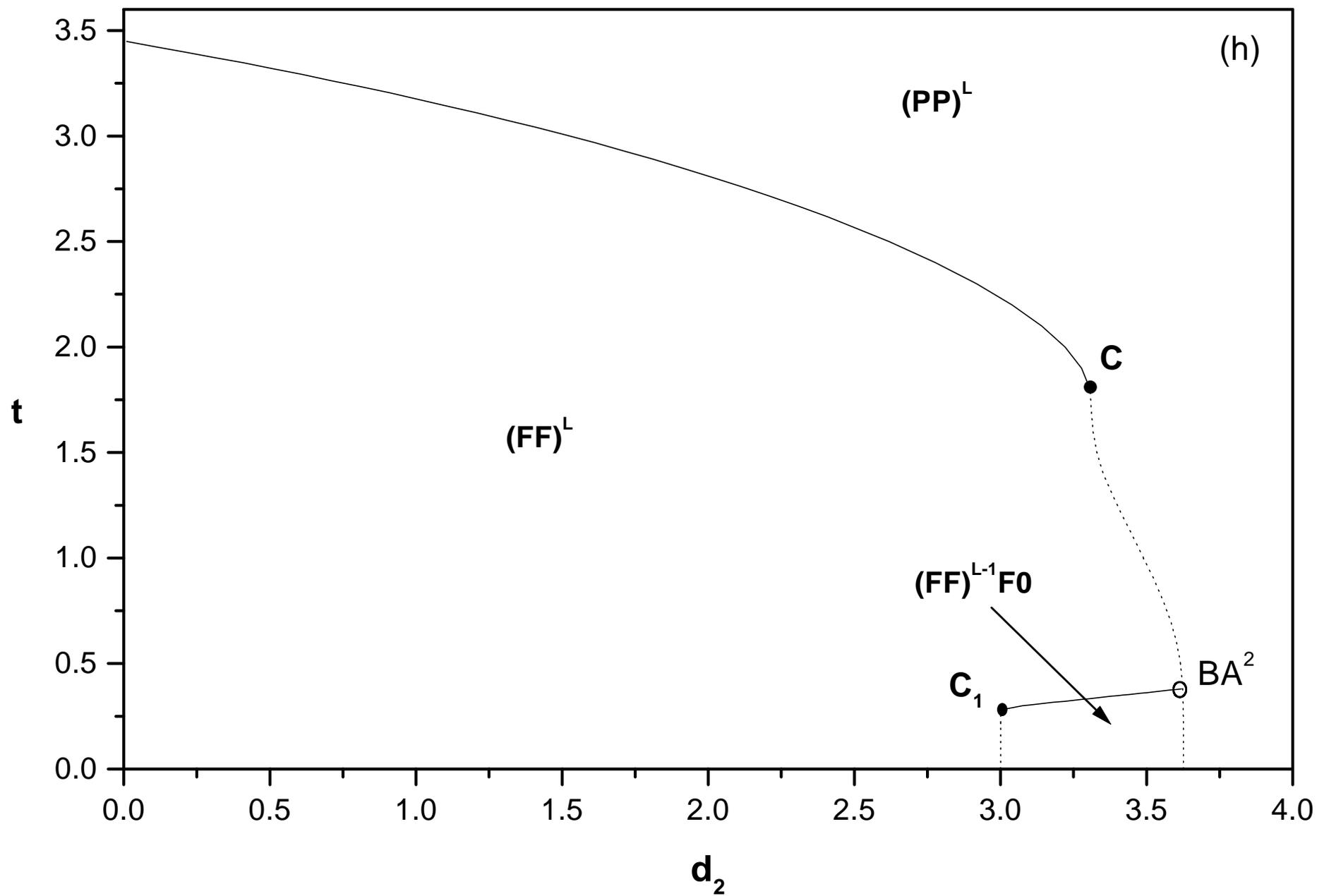
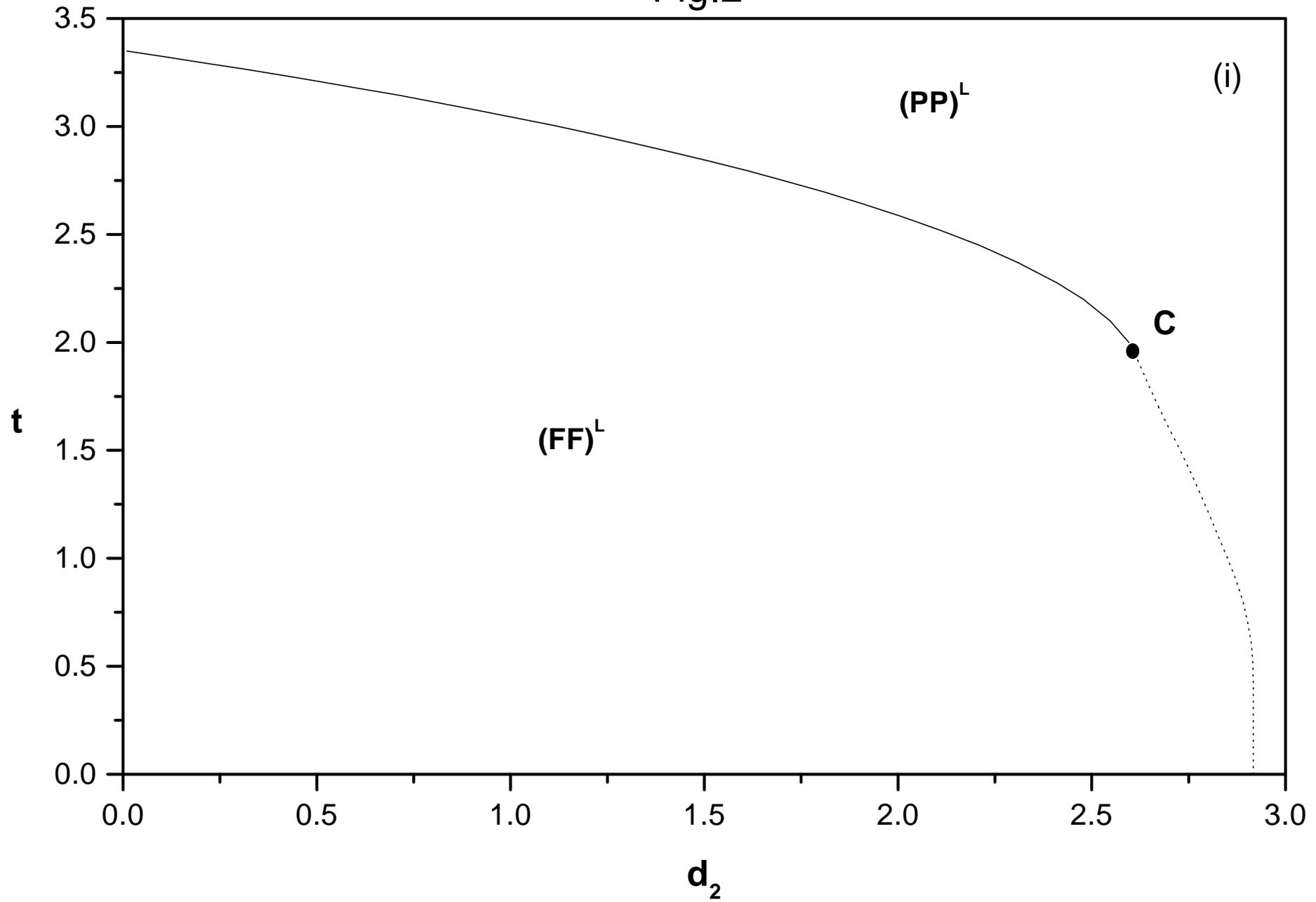
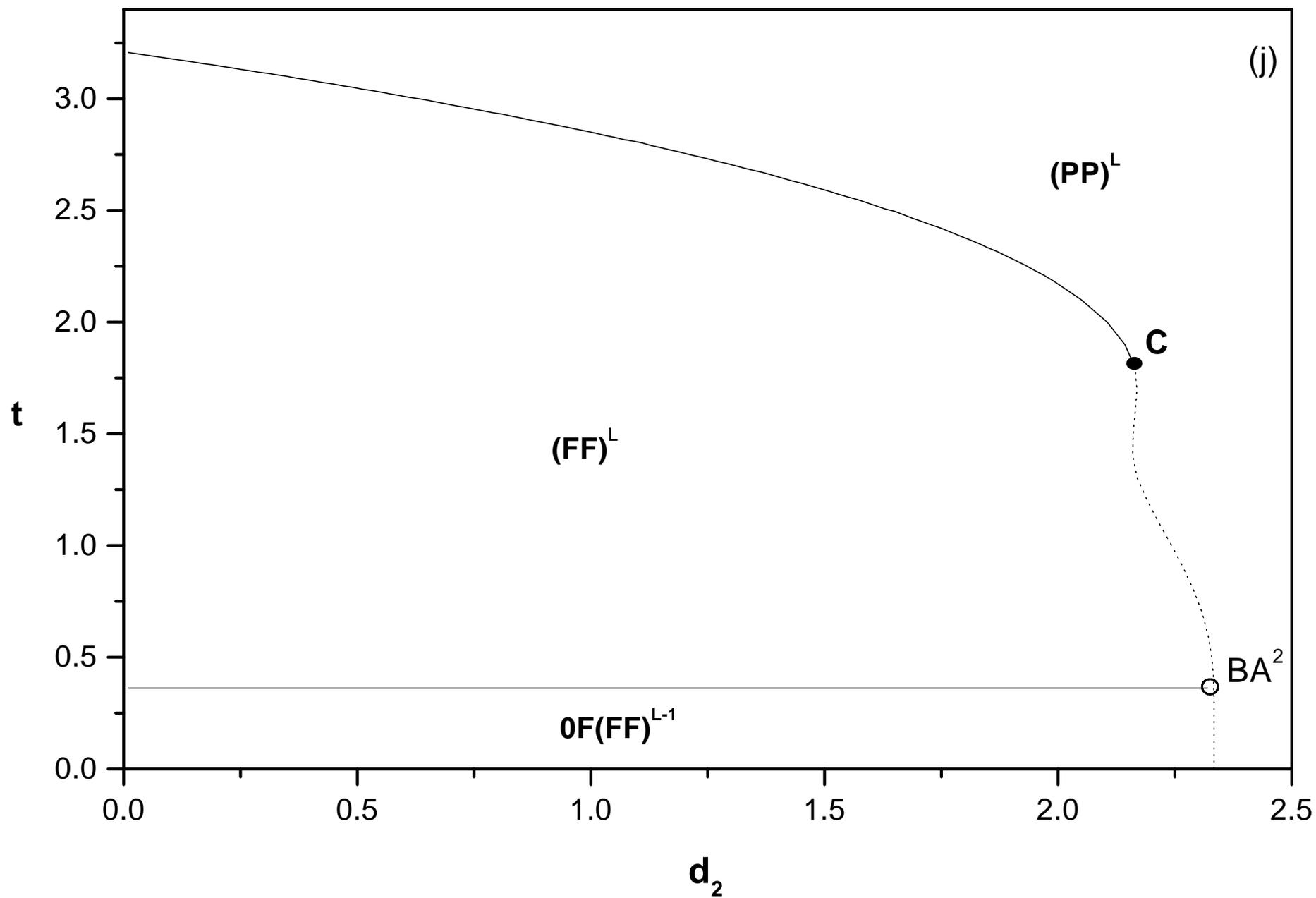
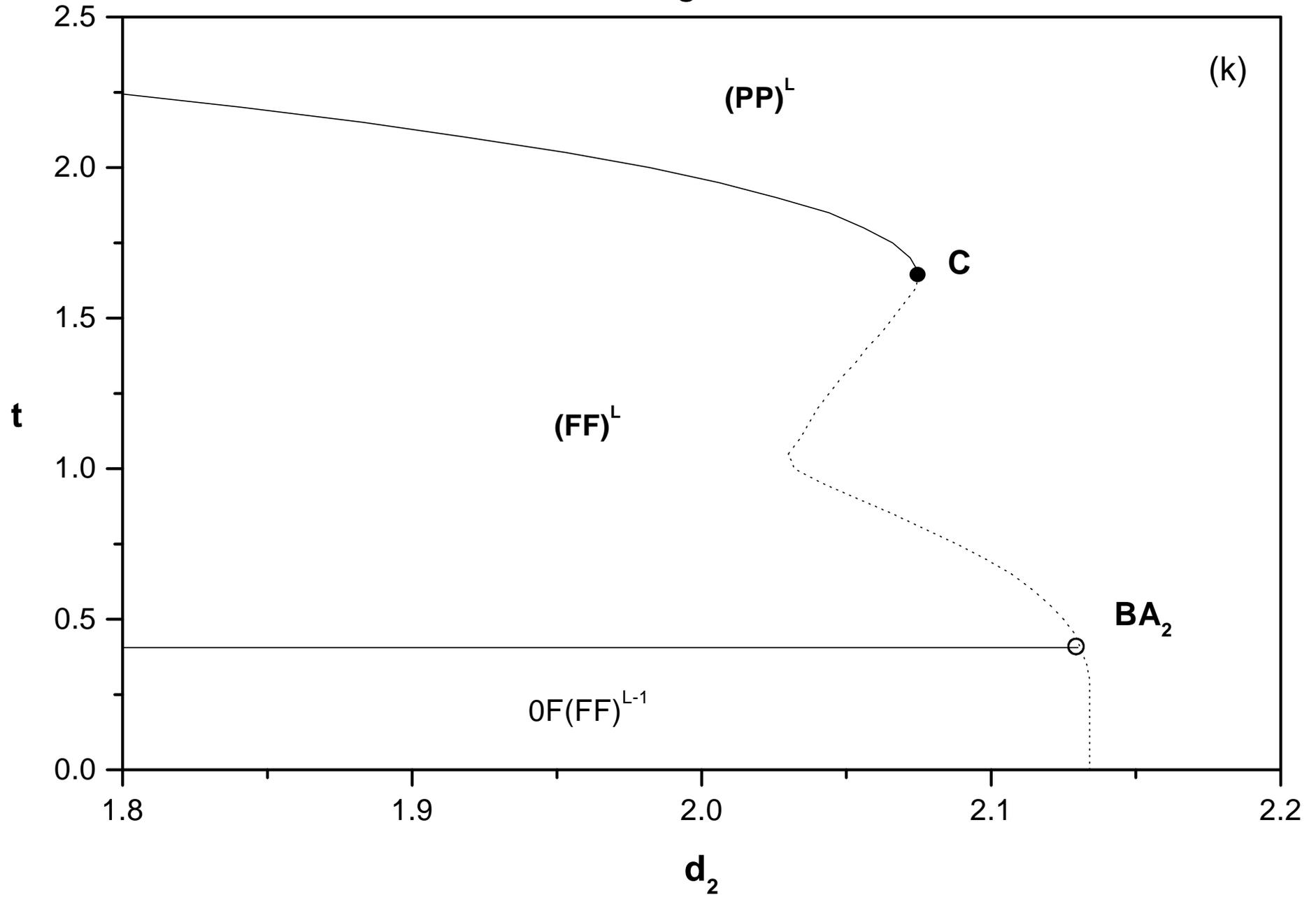
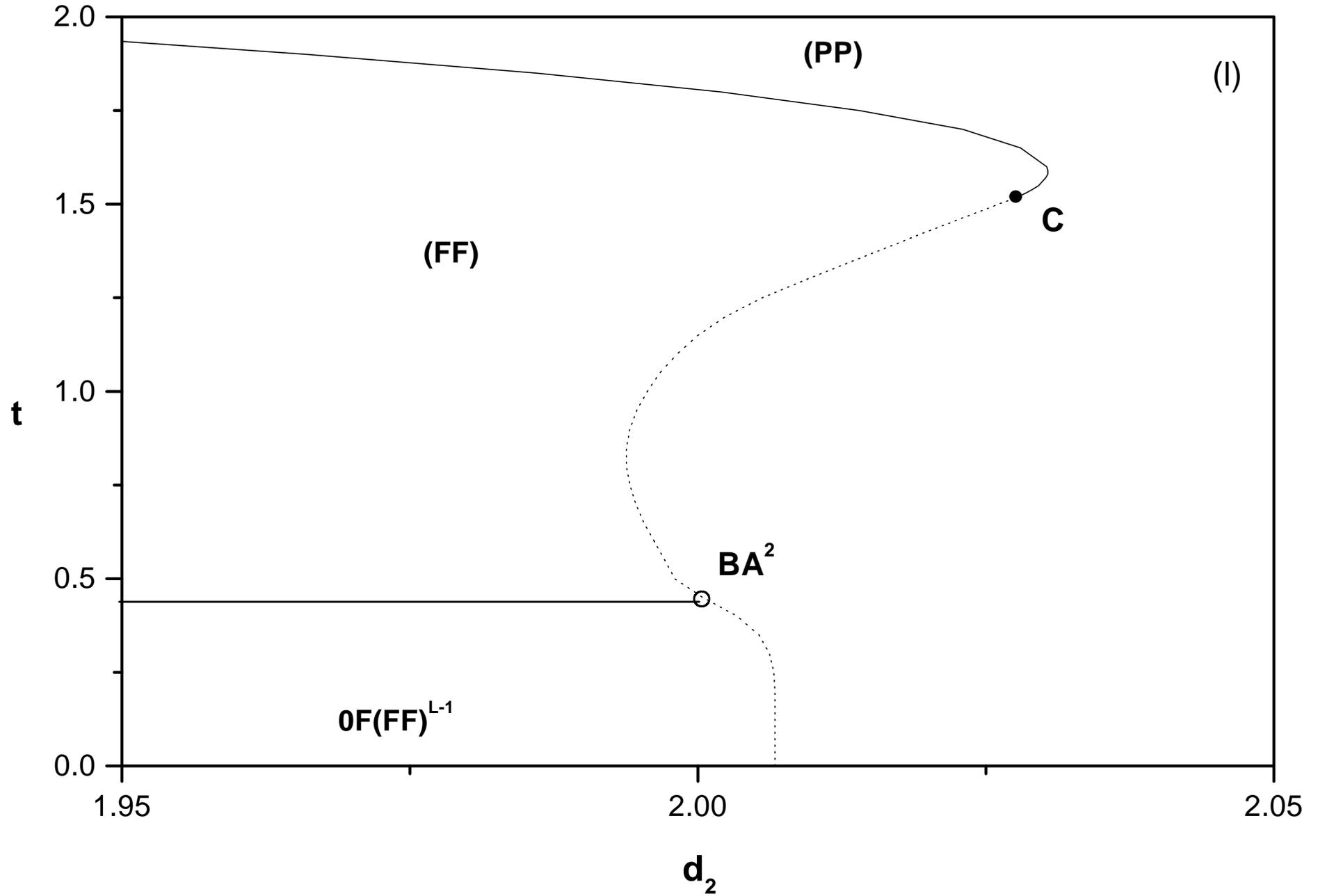
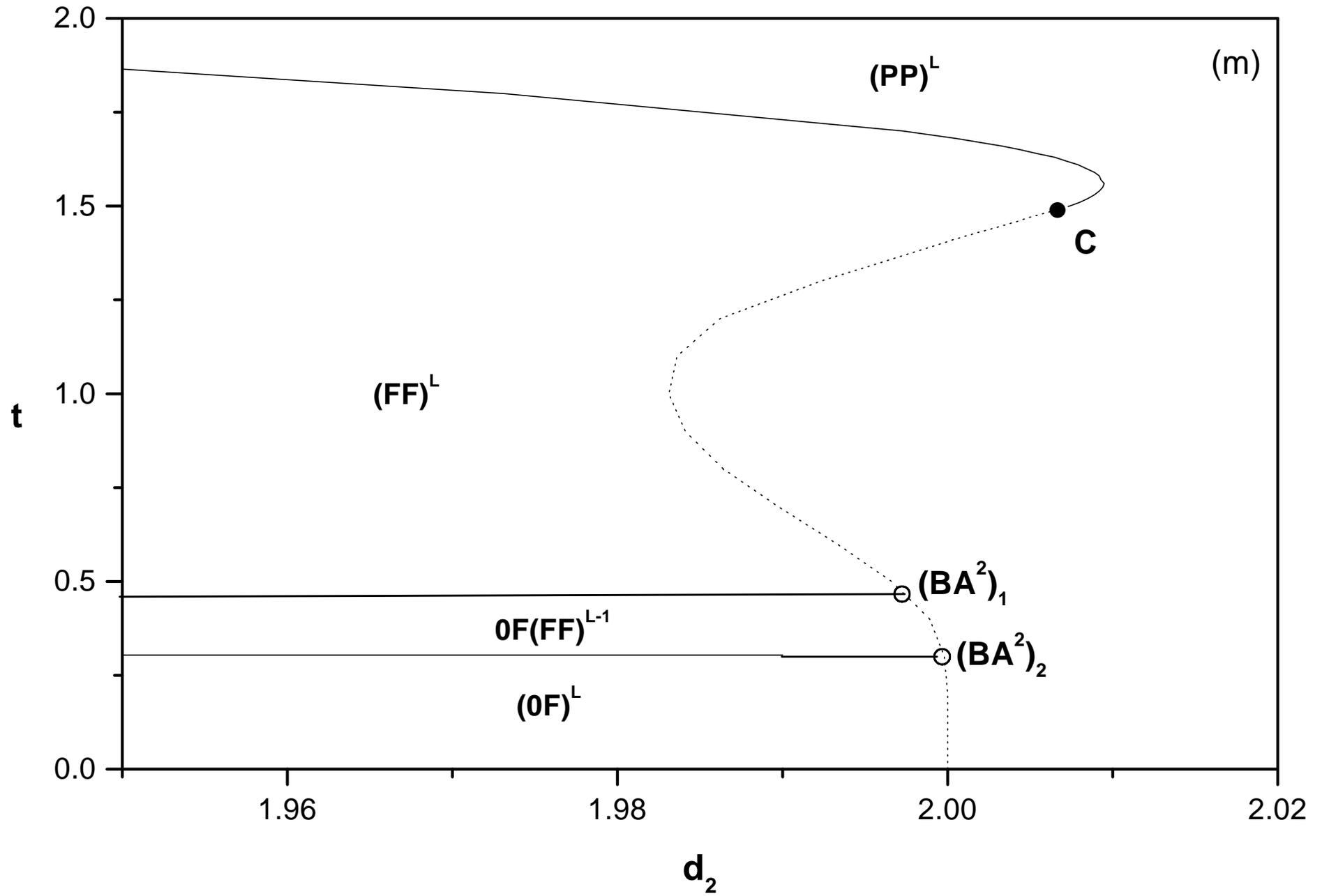
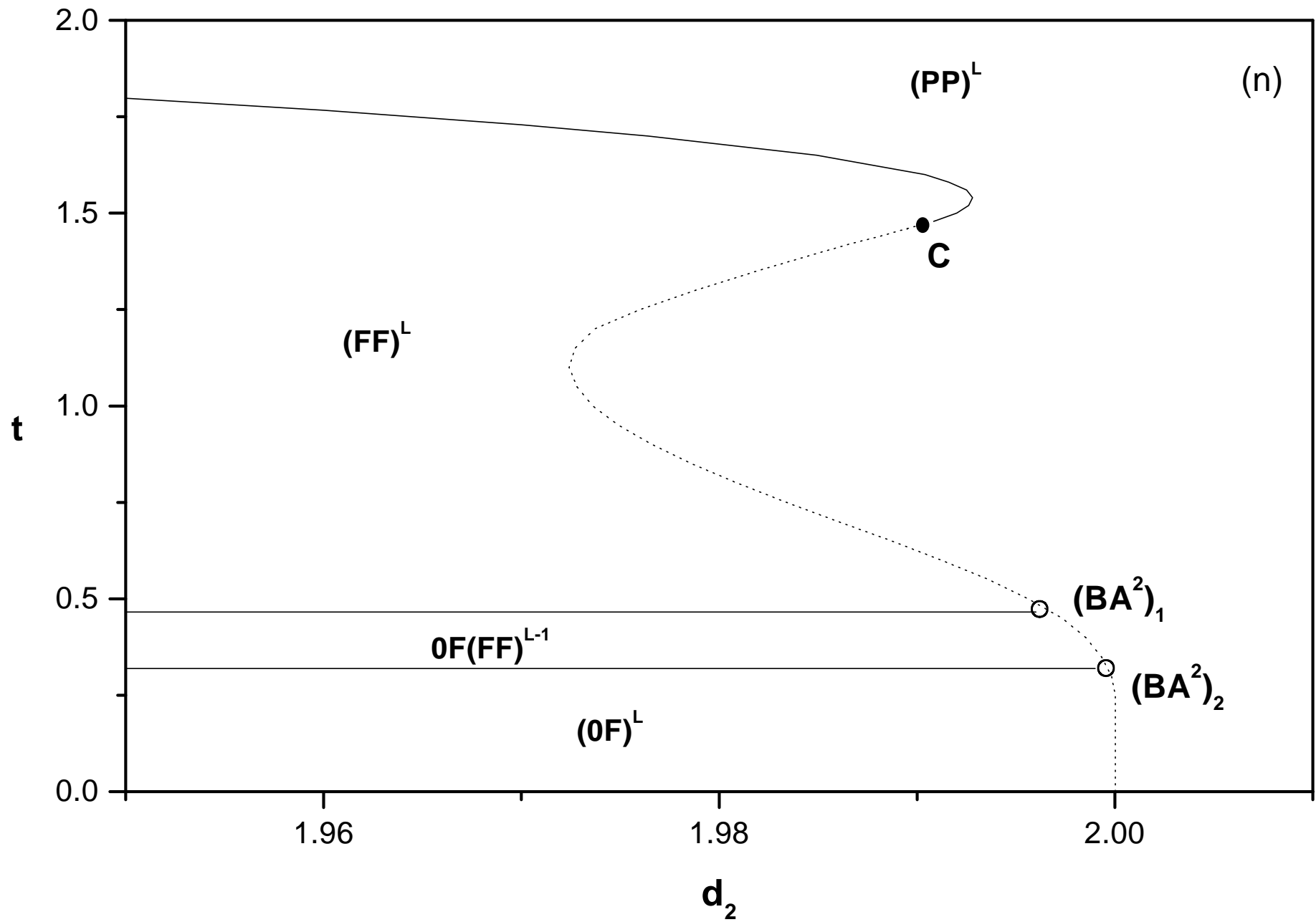
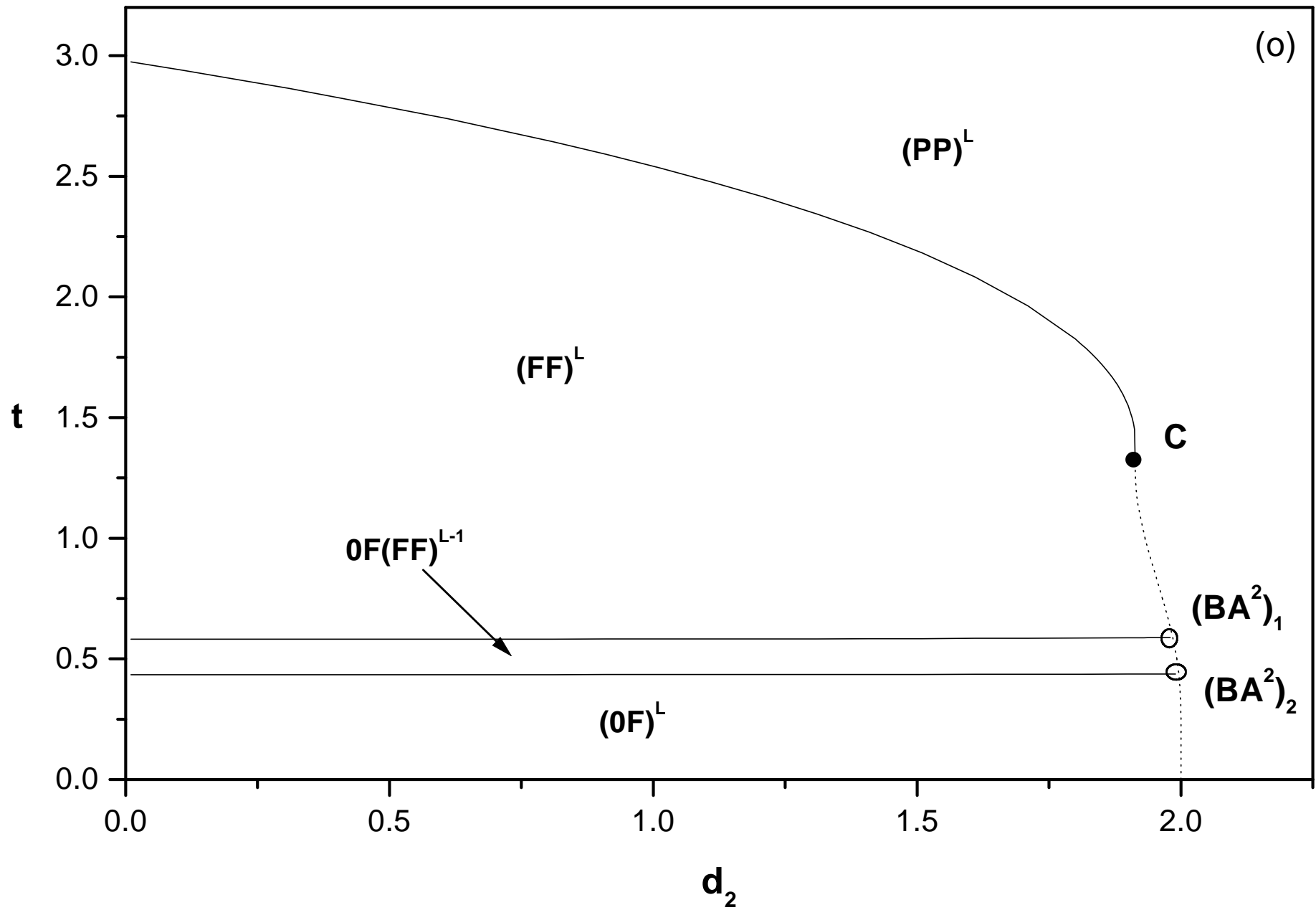
\item[Fig.2] The phase diagrams in the (t,$d_2$) plane for a generic even value of N ($N=6$) and generic values of $d_1$. (a) $d_1=1$., (b)$d_1=1.85$, (c) $d_1=1.855$, (d) $d_1=1.9$, (e) $d_1=1.99$, (f) $d_1=2.01$, (g) $d_1=2.1$, (h) $d_1=2.25$, (i) $d_1=2.75$, (j) $d_1=3.5$, (k) $d_1=3.8$, (l) $d_1=3.9$, (m) $d_1=4.01$, (n) $d_1=4.1$, (o) $d_1=5$. The  solid lines are of second-order, the doted ones are of first-order.
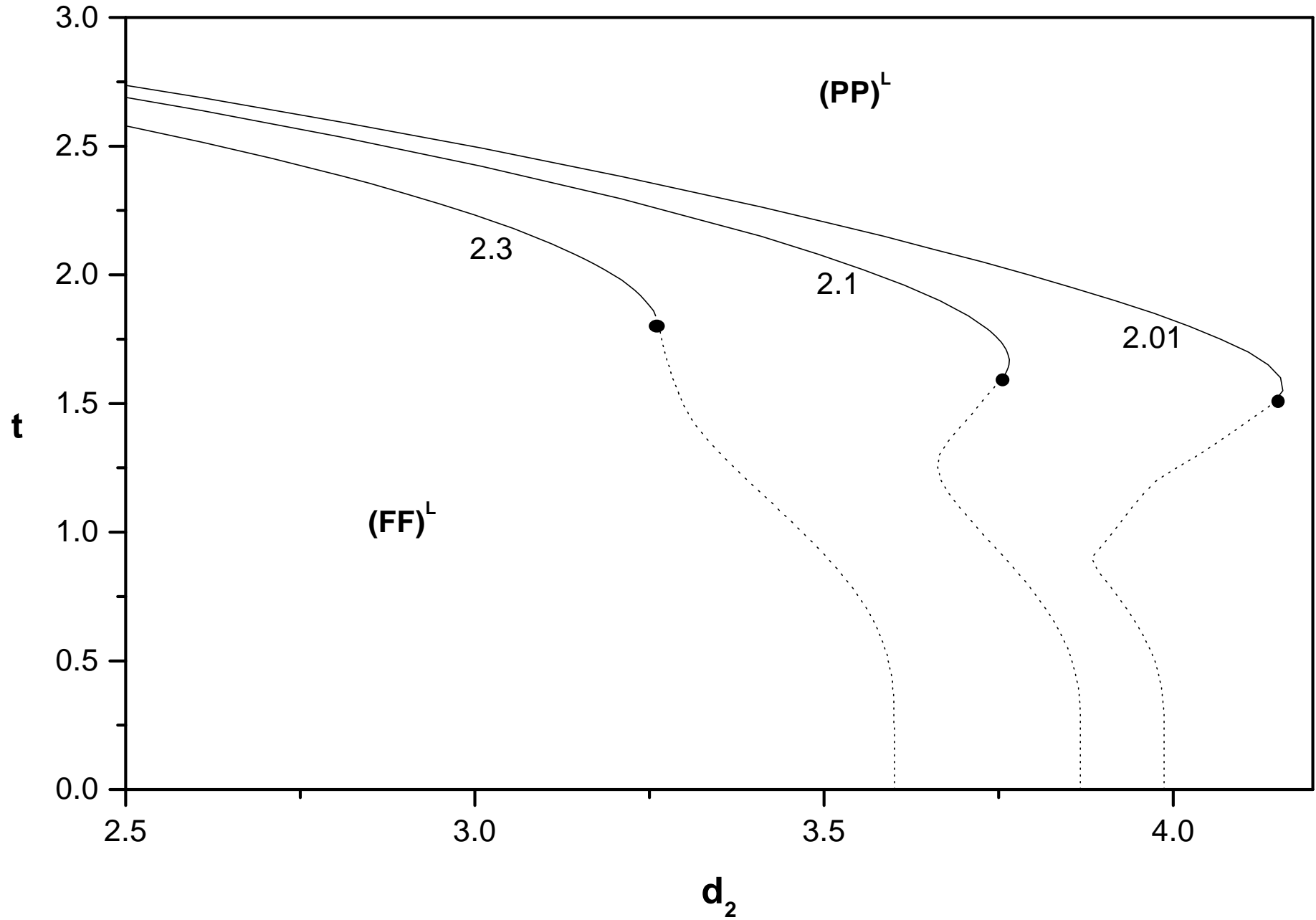
\item[Fig.3] The phase diagram in the (t,$d_2$) plane for a generic odd value of N ($N=7$) and generic values of $d_1$ in the range $2<d_1<3$. The number accompanying each curve is the value of $d_1$. The  solid lines are of second-order, the doted ones are of first-order.  
\end{description}
\end{document}